\documentclass[preprint,reprint,letterpaper,aps,prc,superscriptaddress,nofootinbib,showpacs,floatfix]{revtex4-2}
\usepackage{setspace}
\usepackage{color}
\usepackage{xcolor}
\usepackage{indentfirst}
\usepackage{xspace}
\usepackage{verbatim}
\usepackage{epstopdf}
\usepackage{float}
\usepackage{wrapfig}
\usepackage{graphicx}
\usepackage{amsmath}
\usepackage{amssymb}
\usepackage{amsthm}
\usepackage{natbib}
\usepackage{graphicx}   
\usepackage{amsmath}    
\usepackage{amssymb}    
\usepackage{hyperref}   
\usepackage{natbib}

\begin{document}
\title{Relative transverse activity as a probe of collectivity-like long-range correlations in pp collisions at $\sqrt{s}=13$~TeV}
\author{ Subhadeep~Roy }
\email{subhadeep.roy@cern.ch}
\affiliation{Indian Institute of Technology Bombay, Mumbai 400076, India}

\author{ Sadhana~Dash }
\email{sadhana@phy.iitb.ac.in}
\affiliation{Indian Institute of Technology Bombay, Mumbai 400076, India}

\begin{abstract}
Understanding the origin of collectivity-like signatures in small collision systems is a central open question in high-energy nuclear physics, and two-particle correlation functions offer unique sensitivity to the underlying-event (UE) dynamics that may drive such behavior in proton--proton (pp) collisions.
In this work, the two-particle number ($R_{2}$) and transverse-momentum ($P_{2}$) correlation functions are studied in pp collisions at $\sqrt{s}=13$~TeV using PYTHIA~8, for final-state charged hadrons within $|\eta|<0.8$ and $0.2<p_{\rm T}<2.0$~GeV/$c$, with events classified by the relative transverse activity $R_{\mathrm{T}}$ to probe how UE activity shapes correlation structures in the soft-QCD-dominated regime.
A collectivity-like long-range near-side component is observed in the charge-independent correlator $R_{2}^{\mathrm{CI}}$ exclusively for the highest $R_{\mathrm{T}}$ class ($2.5 < R_{\mathrm{T}} \leq 5.0$), while no corresponding structure appears in the charge-dependent correlators.
This indicates that enhanced UE activity, driven by multiple partonic interactions and color reconnection, can generate collectivity-like long-range correlations without hydrodynamic evolution.
These findings establish $R_{\mathrm{T}}$ as a differential event classifier to provide a non-hydrodynamic baseline for interpreting such signatures in small-system measurements at the LHC.
\end{abstract}
\maketitle

\section{Introduction}
The primary objective of the relativistic heavy-ion collision program is to explore the properties of strongly interacting matter under extreme temperature and energy density, 
where quarks and gluons exist in a deconfined state known as quark-gluon plasma (QGP). Measurements of anisotropic flow and jet quenching provide compelling evidence for the 
formation of such a medium in relativistic heavy-ion collisions \cite{adler2003disappearance,adare2008transverse,star2005,aamodt2010elliptic,adam2016pseudorapidity,aamodt2012harmonic}. 
In particular, the emergence of long-range, ridge-like correlations extending over large pseudorapidity separations ($|\Delta\eta|$) is widely regarded as a hallmark of collective expansion
in these systems. Such long-range correlations are quantitatively described within relativistic hydrodynamic frameworks, and the extracted flow coefficients exhibit characteristic scaling 
patterns consistent with a strongly coupled, collectively expanding medium \cite{adare2007scaling,abelev2015elliptic,adam2016higher}.
\par 
Remarkably, the long-range ridge-like correlation structures have also been observed in high-multiplicity proton--lead (p--Pb) and proton--proton (pp) collisions \cite{CMS:2010ifv, ALICE:2012eyl, ATLAS:2012cix}, which were previously considered too small to sustain the formation of a collectively expanding hydrodynamic medium.
These observations have sparked intense debate regarding the microscopic origin of collectivity-like signatures in small collision systems.
On one hand, they may suggest the emergence of collective dynamics even in systems with limited spatial extent. On the other hand, QCD-driven mechanisms, 
such as multiple partonic interactions (MPI), color reconnection (CR), and gluon saturation effects, are also capable of generating correlation patterns that resemble 
those typically attributed to hydrodynamic flow. Precision observables sensitive to particle production dynamics across different kinematic and event-activity regimes 
are therefore needed to constrain these competing contributions and provide a reliable baseline for interpreting the experimental data.
The two-particle number correlation function, $R_{2}(\Delta \eta, \Delta \varphi)$, and the two-particle transverse momentum 
correlation function, $P_{2}(\Delta \eta, \Delta \varphi)$, are particularly well suited for this purpose \cite{sharma2009methods}. 
The $R_{2}(\Delta \eta, \Delta \varphi)$ probes correlations among particle pairs in terms of multiplicity
fluctuations, while the transverse-momentum correlation function, $P_{2}(\Delta \eta, \Delta \varphi)$, is sensitive to transverse-momentum fluctuations. 
Measurements of $R_{2}$ and $P_{2}$ correlations by the ALICE Collaboration in Pb--Pb collisions at $\sqrt{s_{\rm NN}}=2.76$~TeV \cite{aliceR2P2} revealed pronounced
near- and away-side structures. Notably, $P_{2}$ was found to be consistently narrower than $R_{2}$, reflecting their distinct sensitivities to collective flow, jet correlations,
and local charge conservation \cite{adam2017flow}. Simulation studies with AMPT, EPOS, and UrQMD further demonstrated that the relative contributions of hydrodynamic expansion, 
resonance decays, and jet fragmentation can be disentangled by comparing charge-independent (CI) and charge-dependent (CD) combinations of the $R_{2}$ and $P_{2}$ correlators 
\cite{basu2021differential}. However, while such studies have provided critical information in large systems, the systematic studies of $R_{2}$ and $P_{2}$ correlations in small 
systems such as pp collisions remain far less explored.
\par
In pp collisions, particle production results from both the primary hard partonic scattering and softer QCD processes collectively referred to as the underlying event (UE).
The UE encompasses processes such as MPI, initial- and final-state radiation, beam-beam remnants, and CR, all of which
contribute substantially to the overall event activity. These soft and semi-soft QCD processes dominate particle production at low transverse momentum ($p_{\rm T}$),
the kinematic region where collectivity-like signatures are most likely to manifest. Consequently, correlation measurements performed within well-defined UE samples are
essential for a reliable interpretation of the observed long-range, collectivity-like signatures and for disentangling the underlying physics mechanisms responsible
for their emergence in small collision systems. The relative transverse activity, $R_{\rm T}$, has been proposed as a robust event classifier~\cite{martin2016probing} to
quantify the UE activity in pp collisions. By selecting events according to their $R_{\rm T}$ values, one can systematically investigate how correlation structures
evolve with increasing UE activity, thereby providing insights into the interplay between hard scattering and soft QCD processes in shaping the observed
correlation patterns.

\par
In this work, we present a PYTHIA~8 simulation study of the $R_2$ and $P_2$ correlation functions in pp collisions at $\sqrt{s}=13$~TeV,
with events classified by $R_{\rm T}$. The analysis targets final-state charged hadrons within $|\eta| < 0.8$ and $0.2 < p_{\rm T} < 2.0$~GeV/$c$. 
We examine the possible emergence of long-range structures in the $R_{2}$ and $P_{2}$ correlators
and the evolution of their near-side peak widths with $R_{\rm T}$. 
The results provide a non-hydrodynamic baseline for interpreting collectivity-like signatures in small-system measurements at the LHC and offer insights into the role 
of UE activity in shaping two-particle correlations in pp collisions.

\section{Relative Transverse activity}
\begin{figure}
\centering
\includegraphics[width=0.9\linewidth]{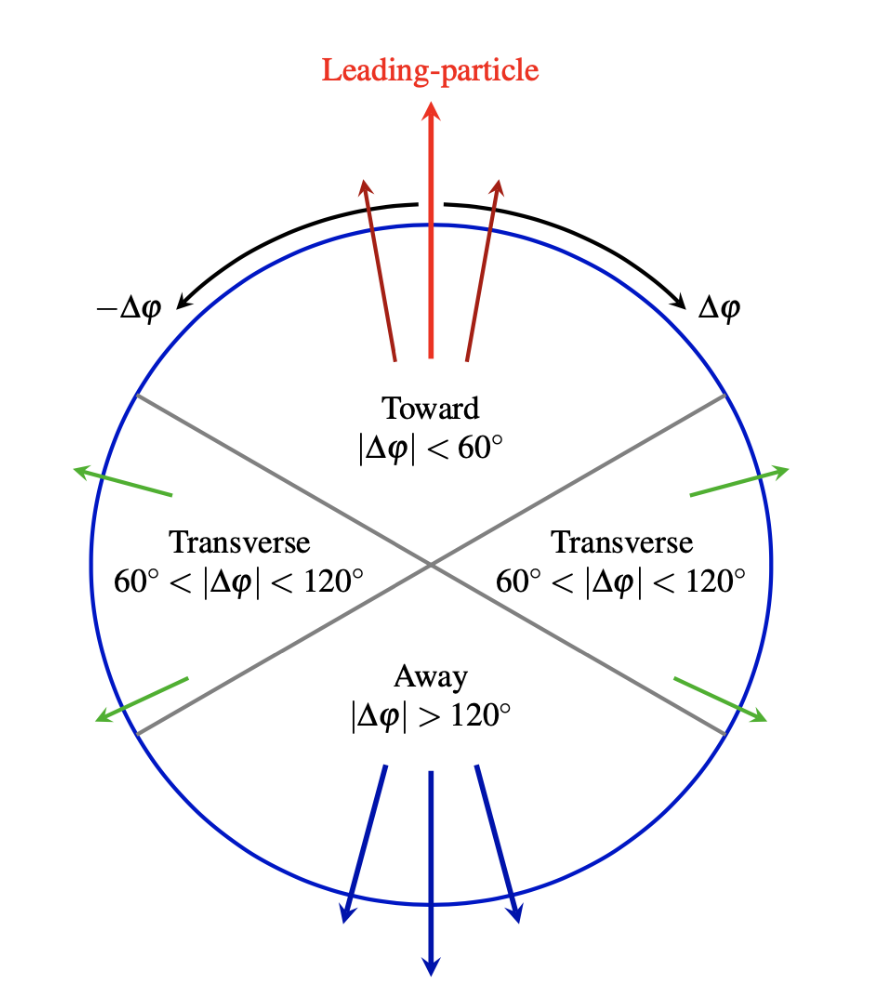}
\caption{Schematic of the event topology used in underlying-event analyses. The azimuthal plane is divided into three regions relative to the leading charged particle: 
the \textit{Toward} ($|\Delta\varphi|<60^{\circ}$) and \textit{Away} ($|\Delta\varphi|>120^{\circ}$) regions, dominated by hard-scattering and jet activity, and the \textit{Transverse} 
region ($60^{\circ}<|\Delta\varphi|<120^{\circ}$), which is most sensitive to the underlying event~\cite{alice2019underlying}.}
\label{topology_dia}
\end{figure}

A standard approach in UE studies is to identify the charged particle carrying the highest transverse momentum ($p_{\rm T}$) in the event,
referred to as the leading particle. The azimuthal plane is then partitioned into three regions defined relative to the azimuthal angle of the leading particle,
$\varphi_{\rm lead}$: the \textit{Toward} region ($|\Delta\varphi|<60^{\circ}$), the \textit{Away} region ($|\Delta\varphi|>120^{\circ}$),
and the \textit{Transverse} region ($60^{\circ}<|\Delta\varphi|<120^{\circ}$), as illustrated in Fig.~\ref{topology_dia}.
The toward and away regions are dominated by hard-scattering products and jet activity, while the transverse region is primarily sensitive to the UE contributions.

Experimental measurements indicate that the charged-particle number density in the transverse region rises steeply with the leading-particle $p_{\rm T}$, reaching a
plateau at $p_{\rm T}^{\rm lead}\gtrsim5$~GeV/$c$ \cite{alice2019underlying}. Above this threshold, the transverse region therefore provides
a clean, hard-scattering-independent probe of the UE activity. In this regime, $R_{\rm T}$ is defined as the ratio of the charged-particle multiplicity in
the transverse region to its minimum-bias average \cite{martin2016probing}:
\begin{equation}
R_{\rm T} = \frac{N_{\rm T}}{\langle N_{\rm T}\rangle},
\end{equation}
where $N_{\rm T}$ is the charged-particle multiplicity in the transverse region for a given event and $\langle N_{\rm T}\rangle$ is its inclusive average over the full event ensemble.
By construction, $R_{\rm T} \ll 1$ selects events with suppressed UE activity in which particle production is predominantly driven by hard-scattering processes,
while $R_{\rm T} \gg 1$ selects events with enhanced UE activity and substantial contributions from soft and semi-soft QCD processes.
\par
\begin{figure}[!htp]
\centering
\includegraphics[width=0.8\linewidth]{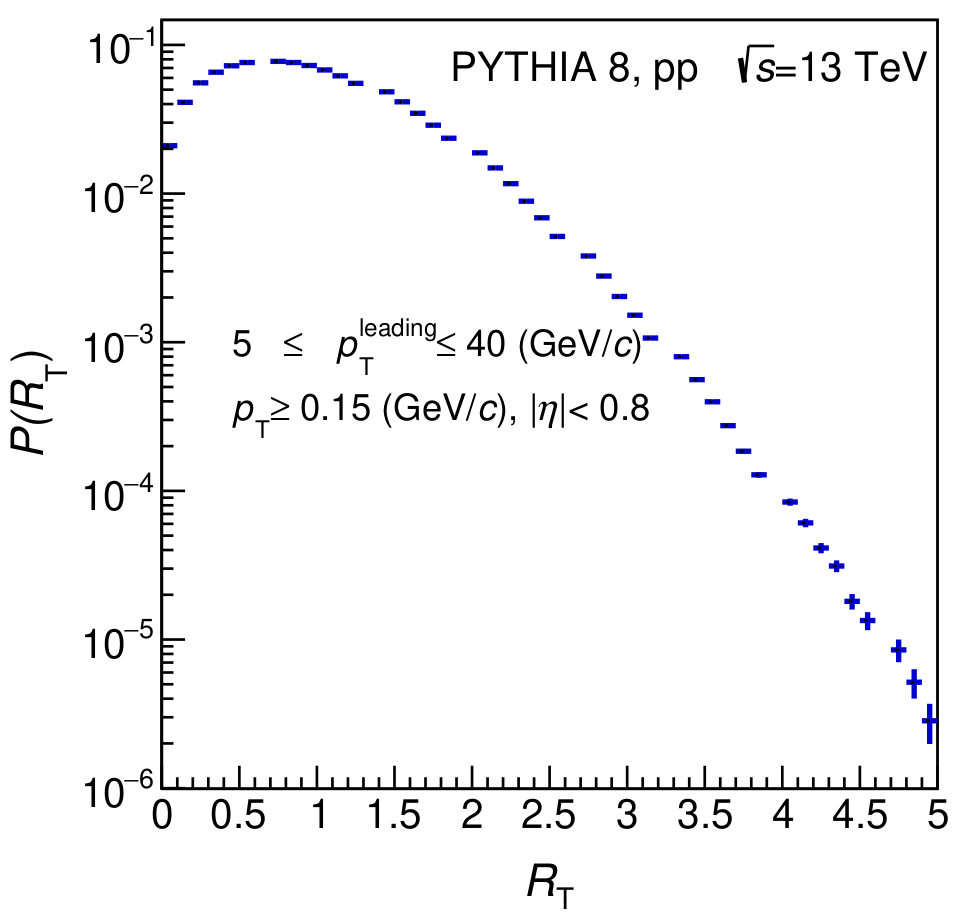}
\caption{Probability distribution of the relative transverse activity $R_{\rm T}$ in pp collisions at $\sqrt{s}=13$~TeV, simulated with PYTHIA~8 (Monash 2013 tune). 
Only events with a leading charged particle in the range $5 \leq p_{\rm T}^{\rm lead} \leq 40$~GeV/$c$ are included, restricting the sample to the transverse-plateau regime where the UE is well-defined. 
The mean transverse-region multiplicity is $\langle N_{\rm T}\rangle = 7.39$.}
\label{Rt_dist}
\end{figure}

 
In this work, $R_{\rm T}$ is computed using charged hadrons within the kinematic acceptance $|\eta| < 0.8$ and $p_{\rm T} \geq 0.15$~GeV/$c$.
Events are further required to contain a leading charged particle satisfying $5 \leq p_{\rm T}^{\rm lead} \leq 40$~GeV/$c$, thereby restricting the analysis to the transverse plateau
region where the UE contribution is well-defined. A total of $5\times10^{8}$ minimum-bias events were generated with PYTHIA~8 using the Monash 2013 tune \cite{MonashTune},
a globally optimized parameter set tuned to LHC data to provide an improved description of soft-QCD and underlying-event observables. Of these, approximately $5\times10^{6}$ events
pass the above selection criteria, ensuring sufficient statistical precision across all $R_{\rm T}$ intervals. The resulting $R_{\rm T}$ distribution for pp collisions at $\sqrt{s}=13$~TeV is shown in Fig.~\ref{Rt_dist}, with a mean transverse charged-particle multiplicity of $\langle N_{\rm T} \rangle = 7.39$.
To suppress potential contamination from wide-angle final-state radiation, the analysis is restricted to $R_{\rm T} \leq 5$.
Within this range, events are categorized into four $R_{\rm T}$ classes: $0 \leq R_{\rm T} \leq 0.5$, $0.5 < R_{\rm T} \leq 1.5$, $1.5 < R_{\rm T} \leq 2.5$, and $2.5 < R_{\rm T} \leq 5.0$.

\begin{figure}[!htp]
 	\centering
 	\includegraphics[width=0.9\linewidth]{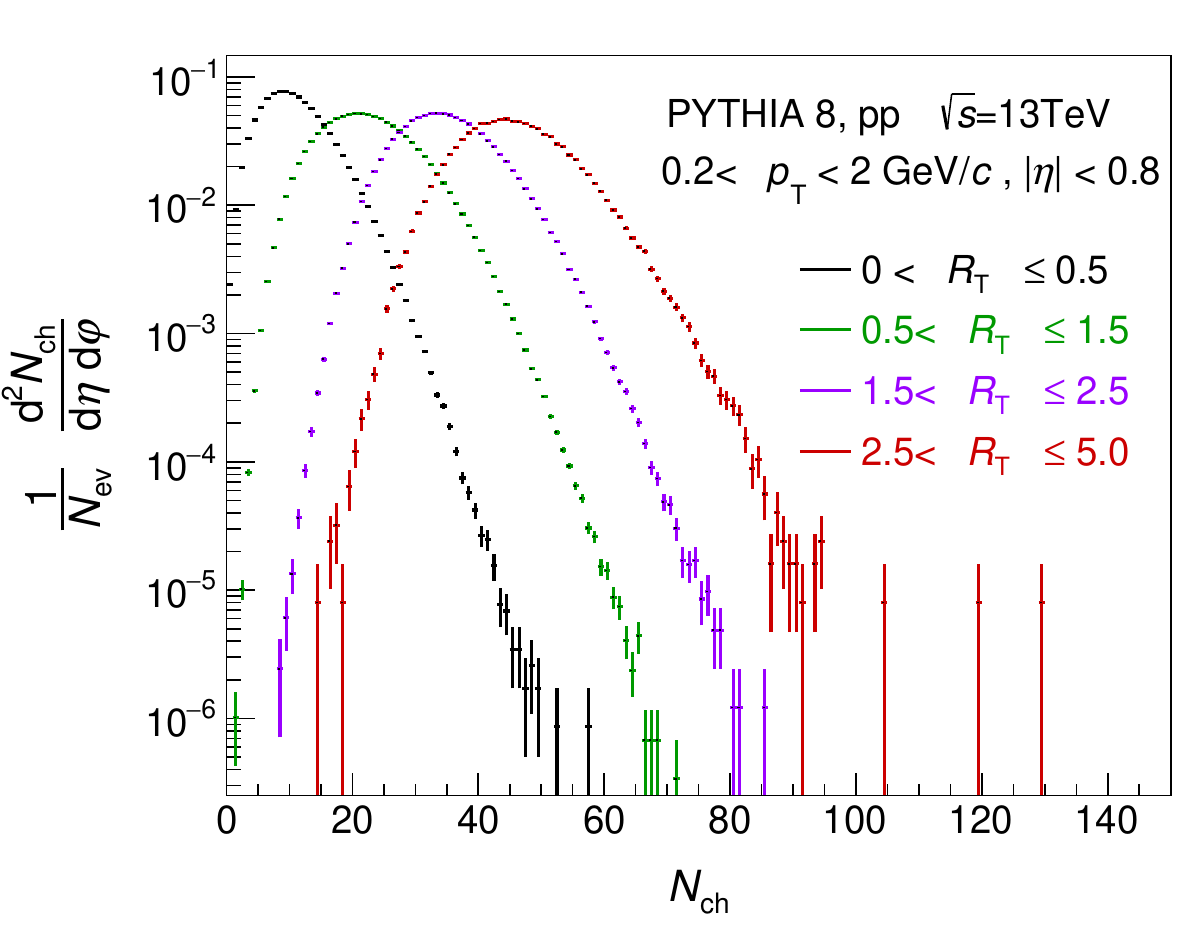}
    \caption{Charged-particle multiplicity ($N_{\rm ch}$) distributions in pp collisions at $\sqrt{s}=13$~TeV simulated with PYTHIA~8, shown separately for each $R_{\rm T}$ interval.
    Charged hadrons within $|\eta|<0.8$ and $0.2 < p_{\rm T} < 2.0$~GeV/$c$ are considered. Events with higher $R_{\rm T}$ display broader distributions shifted toward larger $N_{\rm ch}$,
    with the mean charged-particle multiplicity increasing progressively from low to high $R_{\rm T}$ classes, reflecting the enhanced UE activity in those events.}
 	\label{nch_dist}
 \end{figure}

Figure~\ref{nch_dist} shows the charged-particle multiplicity distributions for events in each $R_{\rm T}$ class.
As expected, events at higher $R_{\rm T}$ exhibit larger overall multiplicities, consistent with an enhanced UE
contribution. Although $R_{\rm T}$ is correlated with the global charged-particle multiplicity, it provides a more differential event classification by isolating the transverse-region
activity relative to the inclusive event average at a fixed leading-particle momentum scale. The $R_{\rm T}$-based selection therefore offers a controlled framework to investigate the evolution
of the two-particle correlation functions $R_{2}$ and $P_{2}$ as a function of UE activity, enabling a more differential disentanglement of soft-QCD and hard-scattering contributions.
\section{Observables} \label{def:ob}
The single- and two-particle densities, which serve as the basis for defining the $R_{2}$ and $P_{2}$ correlators \cite{sharma2009methods}, are given by
\begin{align}
\rho_{1}(\eta_{i}, \varphi_{i}) &= \frac{d^{2}N}{d\eta_{i}\,d\varphi_{i}}, \\
\rho_{2}(\eta_{1}, \varphi_{1}, \eta_{2}, \varphi_{2}) &= \frac{d^{4}N}{d\eta_{1}\,d\varphi_{1}\,d\eta_{2}\,d\varphi_{2}},
\end{align}
where $\eta_{i}$ and $\varphi_{i}$ ($i=1,2$) denote the pseudorapidities and azimuthal angles of particles 1 and 2, respectively.
 \par 
 The $R_{2}$ correlator quantifies particle-number correlations and is defined as the normalized two-particle cumulant:
\begin{align}
R_{2}(\eta_{1}, \varphi_{1}, \eta_{2}, \varphi_{2})
= \frac{\rho_{2}(\eta_{1}, \varphi_{1}, \eta_{2}, \varphi_{2})}
{\rho_{1}(\eta_{1}, \varphi_{1})\,\rho_{1}(\eta_{2}, \varphi_{2})} - 1.
\end{align}
In contrast, the two-particle transverse-momentum correlator, $P_{2}$, characterizes transverse-momentum correlations and is defined as
\begin{align*}
 P_{\rm 2} (\eta_{1}, \varphi_{1}, \eta_{2}, \varphi_{2}) = \frac{\langle \Delta p_{\rm T} \Delta p_{\rm T}\rangle (\eta_{1}, \varphi_{1}, \eta_{2}, \varphi_{2})}{\langle p_{\rm T}\rangle ^2}
 \end{align*}
 \begin{align}
 =\frac{1}{\langle p_{\rm T}\rangle ^2} \frac{\int_{p_{\rm T,min}}^{p_{\rm T,max}} \rho_{2} (\vec{p_{1}}, \vec{p_{2}}) \Delta p_{\rm T,1} \Delta p_{\rm T,2} dp_{\rm T,1} dp_{\rm T,2}}{\int_{p_{\rm T,min}}^{p_{\rm T,max}} \rho_{2} (\vec{p_{1}}, \vec{p_{2}}) dp_{\rm T,1} dp_{\rm T,2}},
 \end{align}
 where $\Delta p_{\rm T,i} = p_{\rm T,i} - \langle p_{\rm T}\rangle$ is the deviation of the $i$-th particle's transverse momentum from the inclusive mean, $\langle p_{\rm T}\rangle = \frac{\int\rho_{1} p_{\rm T}\,dp_{\rm T}}{\int\rho_{1}\,dp_{\rm T}}$, evaluated over the range $p_{\rm T,min} \leq p_{\rm T} \leq p_{\rm T,max}$. By construction, the product $\Delta p_{\rm T,1}\Delta p_{\rm T,2}$ is positive when both particles carry $p_{\rm T}$ either above or below $\langle p_{\rm T}\rangle$, and negative when one particle has $p_{\rm T}>\langle p_{\rm T}\rangle$ and the other $p_{\rm T}<\langle p_{\rm T}\rangle$. Consequently, $P_{2}$ is sensitive to the mean transverse momentum of the correlated particles \cite{aliceR2P2}. Particles produced inside jets typically carry $p_{\rm T}$ values larger than $\langle p_{\rm T}\rangle$, thereby contributing positively to $P_{2}$. Furthermore, jet fragments are subject to approximate angular ordering, whereby the highest-$p_{\rm T}$ particles are emitted at small angles with respect to the jet axis while softer particles are distributed over larger angles. This ordering is reflected in the $\Delta\eta$--$\Delta\varphi$ correlation plane, where higher-$p_{\rm T}$ fragments dominate the near-side peak at small pair separations. As a result, $P_{2}$ tends to exhibit a narrower correlation structure in both $\Delta\eta$ and $\Delta\varphi$ compared to $R_{2}$.
\par 
To quantify the correlation strength as a function of pair separation, we integrate over all remaining coordinates within the fiducial acceptance. The projected correlation functions are then expressed as
\begin{align}
 O (\Delta \eta \Delta \varphi)= \frac{1}{\Omega (\Delta \eta)}\int_{\Omega} O (\Delta\eta_{1}, \bar{\eta}, \Delta\varphi_{1}, \bar{\varphi_{2}}) d\bar{\eta} d\bar{\varphi}.
 \end{align}
Here, $O$ denotes a generic two-particle correlator ($R_{2}$ or $P_{2}$), $\Delta\eta = \eta_{1}-\eta_{2}$ and $\Delta\varphi = \varphi_{1}-\varphi_{2}$ are the pair separations, $\bar{\eta} = (\eta_{1}+\eta_{2})/2$ and $\bar{\varphi} = (\varphi_{1}+\varphi_{2})/2$ are the corresponding pair midpoints, and $\Omega(\Delta\eta)$ is the fiducial-acceptance width in $\bar{\eta}$ at fixed $\Delta\eta$. Throughout this work, the azimuthal difference $\Delta\varphi$ is evaluated in the shifted range $[-\pi/2,\,3\pi/2]$.

The correlation functions are evaluated separately for the four possible charged-particle combinations: $(++)$, $(--)$, $(+-)$, and $(-+)$. The like-sign (LS) and unlike-sign (US) correlations are obtained by averaging over the respective pair categories as
\begin{align}
O^{\rm LS} &= \tfrac{1}{2}\big[(++)+(--)\big], \\
O^{\rm US} &= \tfrac{1}{2}\big[(+-)+(-+)\big].
\end{align}

In high-energy hadronic and nuclear collisions, particle production is governed by conservation laws including energy--momentum, baryon number, strangeness, and electric charge. 
At LHC energies, particle and antiparticle yields are nearly symmetric, so that unlike-sign (US) and like-sign (LS) correlations, considered separately, offer limited sensitivity 
to the underlying production mechanisms. To access the charge-dependence of these correlations, we construct the charge-independent (CI) and charge-dependent (CD) combinations as
\begin{align}
O^{\rm CI} &= \tfrac{1}{2}\big(O^{\rm US} + O^{\rm LS}\big), \label{def:CI}\\
O^{\rm CD} &= \tfrac{1}{2}\big(O^{\rm US} - O^{\rm LS}\big).
\label{def:CD}
\end{align}

The charge-independent correlator $O^{\rm CI}$ quantifies the average correlation strength among all charged-particle pairs, while the charge-dependent correlator $O^{\rm CD}$ isolates the
difference between unlike-sign and like-sign correlations. In particular, the charge-dependent number correlator $R_{2}^{\rm CD}$ can be related to the Balance Function of conserved charges
(electric charge $Q$, baryon number $B$, and strangeness $S$) in the limit where the multiplicities of positively ($q$) and negatively ($\bar{q}$) charged particles are
approximately equal \cite{Bass:2000az, ALICE:2021hjb}. In this limit, $R_{2}^{\rm CD}$ serves as a proxy for the Balance Function and provides sensitivity to charge-balancing dynamics imposed
by conservation laws in both elementary and heavy-ion collisions.

\section{Results and Discussions}
\subsection{Charge-independent (CI) correlation functions}
\begin{figure*} [!htp]
\centering
\includegraphics[width=0.75\linewidth]{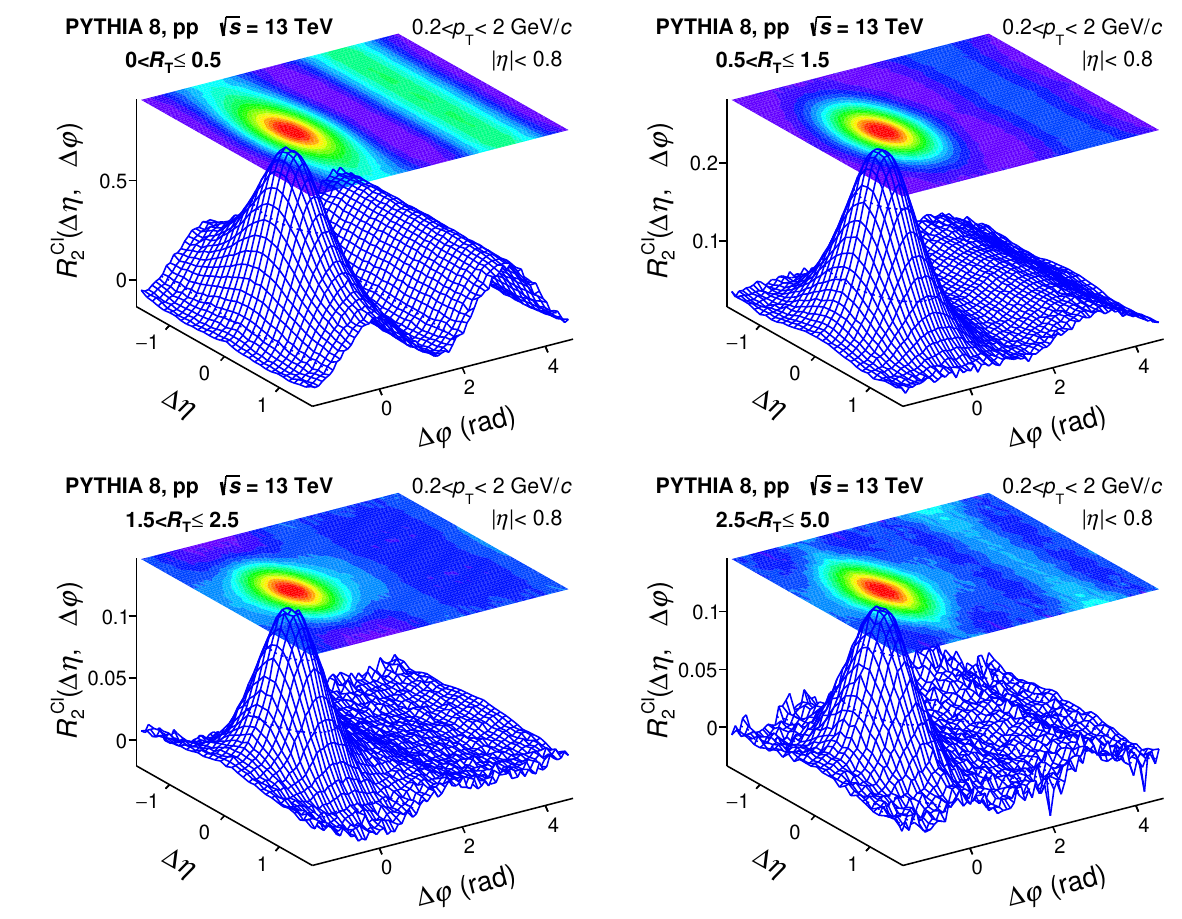}
\caption{Charge-independent two-particle number correlation function $R_{2}^{\rm CI}(\Delta\eta,\Delta\varphi)$ in pp collisions at $\sqrt{s}=13$~TeV simulated with PYTHIA~8 
(Monash 2013 tune), for four $R_{\rm T}$ event classes. Charged hadrons within $|\eta|<0.8$ and $0.2 < p_{\rm T} < 2.0$~GeV/$c$ are considered.}
\label{fig:r2ci_RT4}
\end{figure*}

\begin{figure*}[!htp]
\centering
\includegraphics[width=0.75\linewidth]{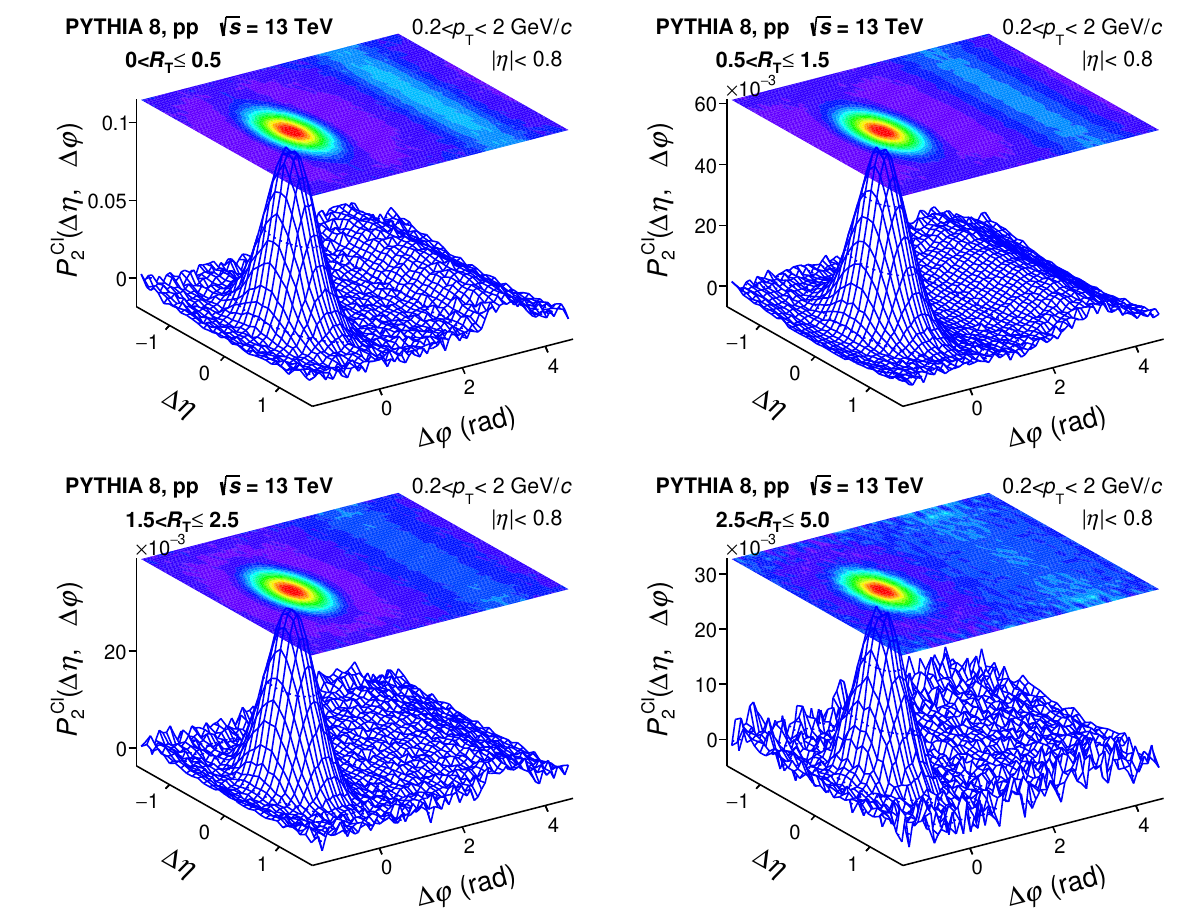}
\caption{Charge-independent two-particle transverse-momentum correlation function $P_{2}^{\rm CI}(\Delta\eta,\Delta\varphi)$ in pp collisions at $\sqrt{s}=13$~TeV simulated with PYTHIA~8 
(Monash 2013 tune), for four $R_{\rm T}$ event classes. Charged hadrons within $|\eta|<0.8$ and $0.2 < p_{\rm T} < 2.0$~GeV/$c$ are considered.}
\label{fig:p2ci_RT4}
\end{figure*}

Figures~\ref{fig:r2ci_RT4} and \ref{fig:p2ci_RT4} present the charge-independent 
two-particle number ($R_{2}^{\rm CI}$) and transverse-momentum ($P_{2}^{\rm CI}$) 
correlation functions for different $R_{\rm T}$ classes, constructed from 
charged particles in PYTHIA~8-generated pp events at $\sqrt{s}=13$~TeV within 
$|\eta|<0.8$ and $0.2<p_{\rm T}<2.0$~GeV/$c$.

Both $R_{2}^{\mathrm{CI}}$ and $P_{2}^{\mathrm{CI}}$ display a prominent near-side
peak at $(\Delta\eta,\Delta\varphi) = (0,0)$ across all $R_{\mathrm{T}}$ classes,
arising from short-range correlations due to jet fragmentation, resonance decays,
and Bose--Einstein effects. A broad away-side ridge at $\Delta\varphi \approx \pi$,
extending across the full $\Delta\eta$ range, is present in both correlators,
reflecting global transverse-momentum conservation and back-to-back dijet topology.
The jet-dominated class ($0 < R_{\mathrm{T}} \leq 0.5$) exhibits the most pronounced
near- and away-side peaks, with $R_{2}^{\mathrm{CI}}$ further displaying a
characteristic $\Delta\eta$-elongated near-side structure consistent with
hard-scattering dominance and the extended pseudorapidity range over which jet
fragments are correlated. With increasing $R_{\mathrm{T}}$, both structures
progressively weaken and broaden, indicating that growing UE activity dilutes
the jet-induced signal.

Notably, $P_{2}^{\mathrm{CI}}$ exhibits systematically narrower near-side peaks 
in both $\Delta\eta$ and $\Delta\varphi$ compared to $R_{2}^{\mathrm{CI}}$ across 
all $R_{\mathrm{T}}$ classes, in agreement with prior ALICE measurements 
\cite{ALICEr2p2pp}. This behavior reflects the sensitivity of $P_{2}$ to 
per-pair transverse-momentum fluctuations 
$\Delta p_{\mathrm{T},1}\,\Delta p_{\mathrm{T},2}$, which suppresses contributions 
from soft pairs whose momenta lie near $\langle p_{\mathrm{T}} \rangle$. 
Consequently, $P_{2}^{\mathrm{CI}}$ is predominantly sensitive to harder, more 
collimated particle pairs, yielding narrower correlation peaks. In contrast, 
$R_{2}^{\mathrm{CI}}$ weights all pairs equally and therefore retains significant 
contributions from soft, isotropic underlying-event particles, resulting in broader 
correlation structures.\par

\begin{figure*} [!htp]
\centering
\includegraphics[width=0.8\linewidth]{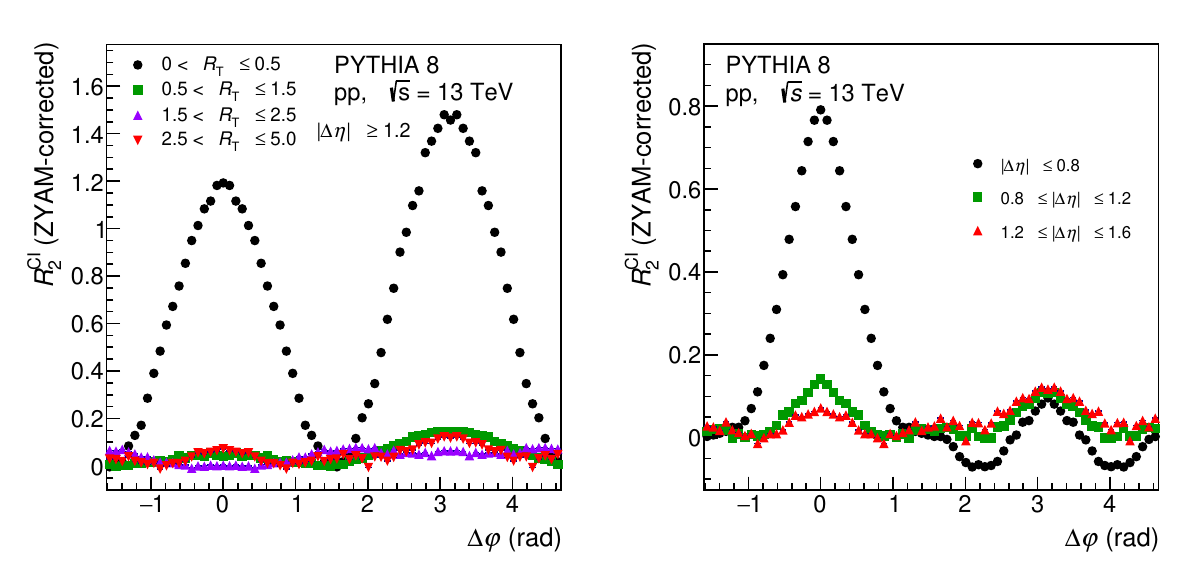}
\caption{ZYAM-corrected $\Delta\varphi$ projections of the charge-independent two-particle number correlation function $R_{2}^{\rm CI}$ in pp collisions at $\sqrt{s}=13$~TeV simulated with PYTHIA~8 (Monash 2013 tune).
(Left) Projections integrated over $1.2 \leq |\Delta\eta| \leq 1.6$, shown for all four $R_{\rm T}$ event classes. A finite near-side excess persists in the highest $R_{\rm T}$ class ($2.5 < R_{\rm T} \leq 5.0$), indicating a long-range ridge-like structure driven by enhanced UE activity.
(Right) Projections for the highest $R_{\rm T}$ class in three $|\Delta\eta|$ intervals: $|\Delta\eta| \leq 0.8$ (jet-core region), $0.8 \leq |\Delta\eta| \leq 1.2$ (transition region), and $1.2 \leq |\Delta\eta| \leq 1.6$ (long-range region). The survival of the near-side excess in the outermost interval confirms that the ridge extends beyond the jet-fragmentation region.}
\label{fig:r2ci_proj_rt_eta}
\end{figure*}

The most striking feature of the $R_{2}^{\rm CI}$ correlations is the emergence 
of a finite long-range near-side component in the UE-dominated class, 
$2.5 < R_{\rm T} \leq 5.0$. This is illustrated in Fig.~\ref{fig:r2ci_proj_rt_eta} 
(left), which shows the ZYAM (zero-yield at minimum)-corrected $\Delta\varphi$ projections of 
$R_{2}^{\rm CI}$ at large pseudorapidity separation, $1.2 \leq |\Delta\eta| \leq 1.6$, 
for all $R_{\rm T}$ classes. In the ZYAM procedure, a polynomial baseline is 
fitted to the near-side region ($|\Delta\varphi| \leq 1.6$) and the minimum of 
the fit is subtracted, isolating the residual near-side yield above the pedestal. 
A clear near-side excess persists at $1.2 \leq |\Delta\eta| \leq 1.6$ in the 
highest $R_{\rm T}$ class, whereas the two intermediate classes, 
$0.5 < R_{\rm T} \leq 1.5$ and $1.5 < R_{\rm T} \leq 2.5$, show a substantially 
reduced or absent long-range component. The lowest-activity class 
($0 < R_{\rm T} \leq 0.5$) also exhibits sizable correlations at large $|\Delta\eta|$; 
however, these arise from hard scattering and jet fragmentation, which can
correlate particle pairs across broad pseudorapidity ranges, and therefore
have a different dynamical origin from the ridge observed in the highest $R_{\rm T}$
class. The persistence of a finite near-side yield in the UE-rich class suggests 
that $R_{\rm T}$-based event selection effectively isolates long-range correlation 
structures that are distinct from jet-induced contributions.

This conclusion is reinforced by the $\Delta\varphi$ projections of $R_{2}^{\rm CI}$ 
in three $|\Delta\eta|$ intervals for the highest $R_{\rm T}$ class, shown in 
Fig.~\ref{fig:r2ci_proj_rt_eta} (right): $|\Delta\eta| \leq 0.8$ (jet core), 
$0.8 \leq |\Delta\eta| \leq 1.2$ (transition region), and 
$1.2 \leq |\Delta\eta| \leq 1.6$ (long-range region). The survival of a near-side 
excess in the outermost interval confirms that the ridge extends well beyond the 
jet-fragmentation region, ruling out a residual near-side jet contribution as its 
source.

Within the PYTHIA~8 framework, this long-range structure originates from the
interplay of MPI and CR,
which generate collective-like azimuthal coherence at the hadronization stage
and produce azimuthally collimated particle yields over extended $\Delta\eta$ ranges.
Its confinement to the highest $R_{\rm T}$ class suggests that sufficiently enhanced
UE activity, mediated through MPI and CR, may give rise to ridge-like long-range
azimuthal correlations in small-system collisions without the need for hydrodynamic
collectivity.

\subsection{Charge-dependent (CD) correlation functions}
\begin{figure*} [!htp]
\centering
\includegraphics[width=0.75\linewidth]{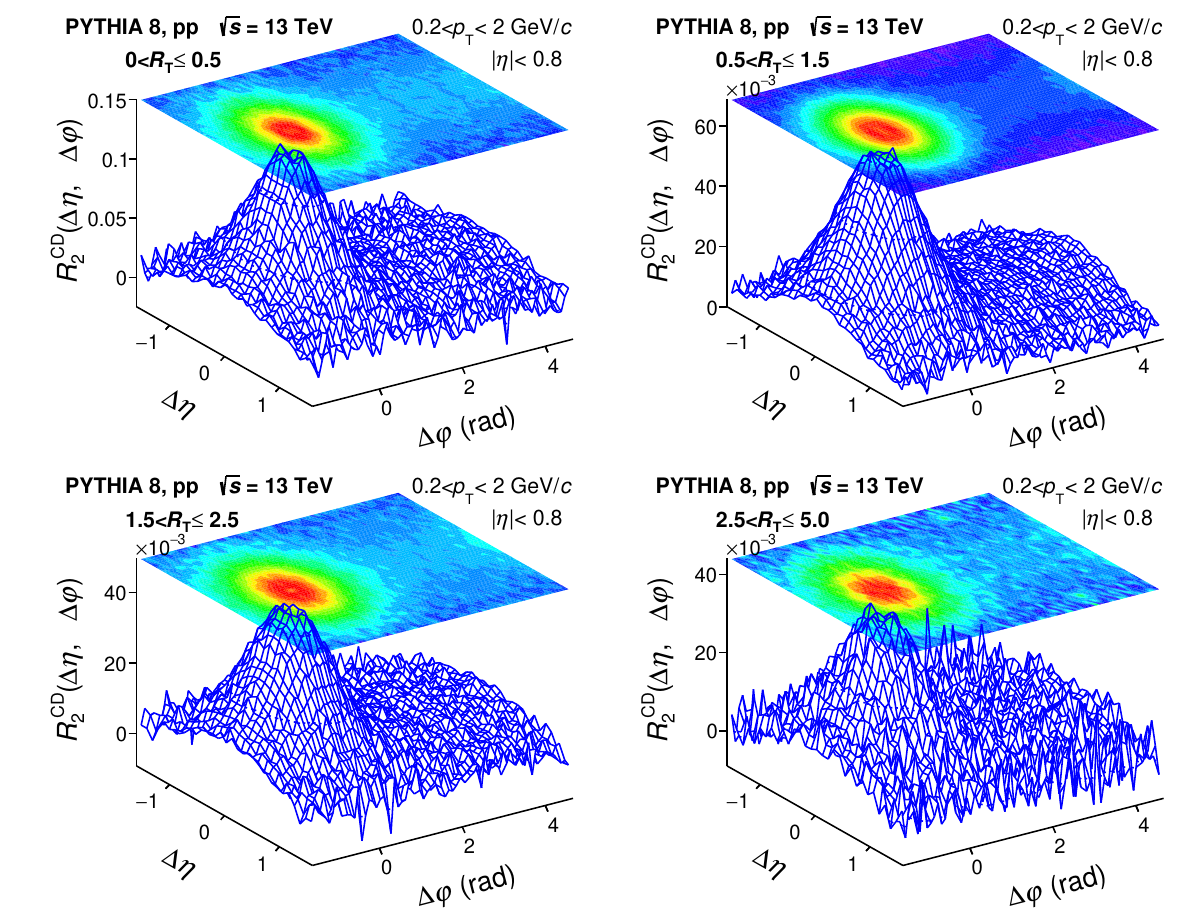}
\caption{Charge-dependent two-particle number correlation function $R_{2}^{\rm CD}(\Delta\eta,\Delta\varphi)$ in pp collisions at $\sqrt{s}=13$~TeV simulated with PYTHIA~8 (Monash 2013 tune), 
for four $R_{\rm T}$ event classes. Charged hadrons within $|\eta|<0.8$ and $0.2 < p_{\rm T} < 2.0$~GeV/$c$ are considered.}
\label{fig:r2cd_RT4}
\end{figure*}

\begin{figure*} [!htp]
\centering
\includegraphics[width=0.75\linewidth]{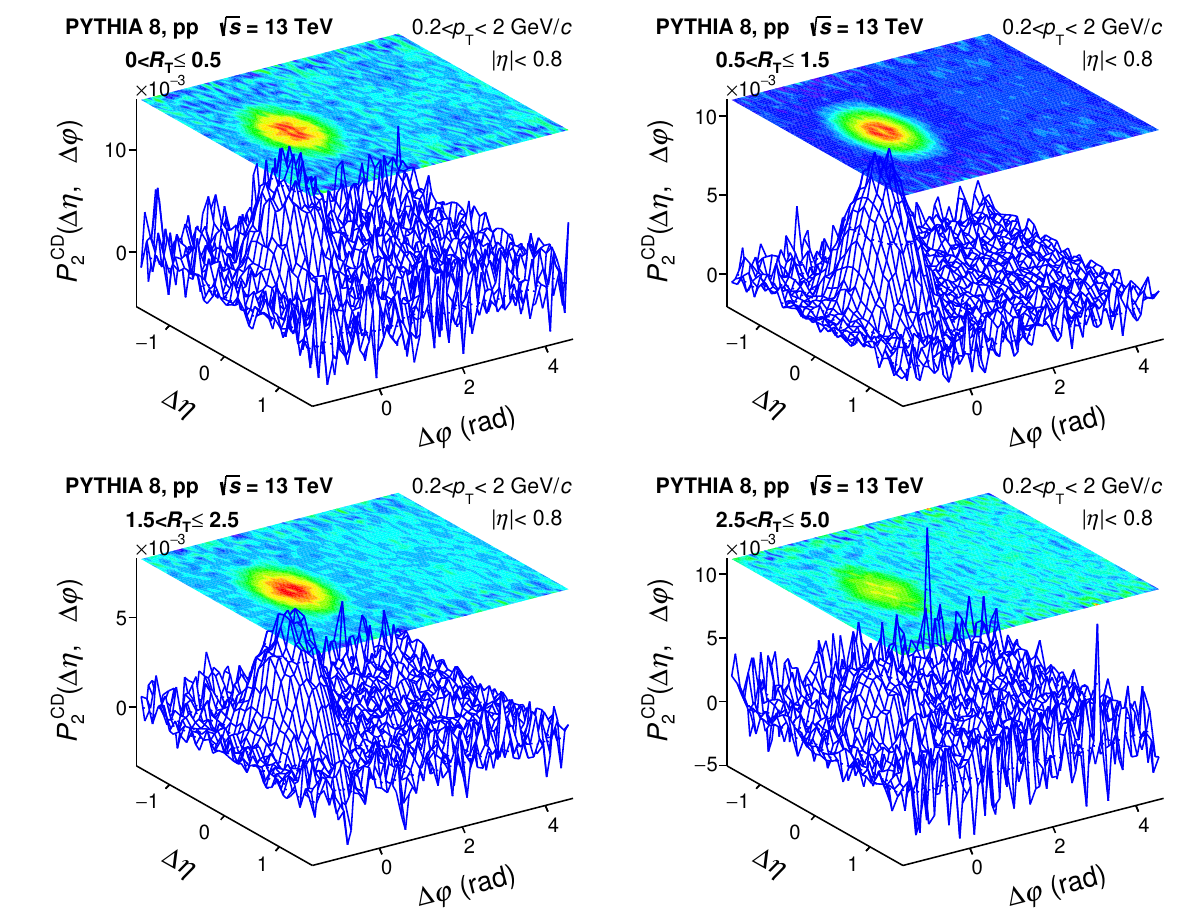}
\caption{Charge-dependent two-particle transverse-momentum correlation function $P_{2}^{\rm CD}(\Delta\eta,\Delta\varphi)$ in pp collisions at $\sqrt{s}=13$~TeV simulated with PYTHIA~8 
(Monash 2013 tune), for four $R_{\rm T}$ event classes. Charged hadrons within $|\eta|<0.8$ and $0.2 < p_{\rm T} < 2.0$~GeV/$c$ are considered.}
\label{fig:p2cd_RT4}
\end{figure*}

Figures~\ref{fig:r2cd_RT4} and \ref{fig:p2cd_RT4} show the charge-dependent
two-particle number ($R_{2}^{\rm CD}$) and transverse-momentum ($P_{2}^{\rm CD}$)
correlation functions for the selected $R_{\mathrm{T}}$ classes.
By construction, the CD correlators isolate the difference between unlike-sign and
like-sign pairs, rendering them primarily sensitive to local charge-balancing
dynamics driven by conservation of electric charge, baryon number, and strangeness.
Within the PYTHIA~8 framework, the CD correlators are therefore directly sensitive
to the production of correlated charge-anticharge pairs in string fragmentation
and their angular separation at hadronization.

Both $R_{2}^{\rm CD}$ and $P_{2}^{\rm CD}$ exhibit a single, localized near-side
peak centered at $(\Delta\eta,\Delta\varphi)=(0,0)$, while the away-side region
remains largely structureless across all $R_{\rm T}$ classes. In contrast to the
CI case, no significant long-range near-side component is observed at large
$|\Delta\eta|$ in any $R_{\rm T}$ interval, indicating that charge balancing
occurs predominantly over short pseudorapidity ranges, consistent with the local
string fragmentation mechanism in PYTHIA~8. Although back-to-back jet production
can generate particle-pair correlations over wide angular separations, charge
conservation within individual parton showers largely confines the balancing charge
to the vicinity of its production point, yielding the observed localized near-side
structure.

As observed for the CI correlators, $P_{2}^{\rm CD}$ exhibits systematically
narrower peaks than $R_{2}^{\rm CD}$, reflecting its sensitivity to
transverse-momentum fluctuations, which enhances contributions from more collimated,
higher-$p_{\rm T}$ particle pairs. With increasing $R_{\rm T}$, both CD correlators
display a gradual reduction in peak magnitude, consistent with a dilution of locally
balanced charge pairs by uncorrelated soft particles from multiple partonic
interactions. Notably, unlike the CI case, neither $R_{2}^{\rm CD}$ nor
$P_{2}^{\rm CD}$ develops a long-range near-side component even in the highest
$R_{\rm T}$ class.

The absence of a long-range structure in the CD correlators, even in the
UE-dominated class, indicates that the ridge-like component observed in
$R_{2}^{\rm CI}$ is not driven by local charge conservation. Rather, it arises
from charge-independent mechanisms, primarily MPI and color reconnection, which
generate azimuthal coherence irrespective of the charge sign of the participating
particles. The contrasting behaviors of the CI and CD correlators thus suggest that
long-range structures in PYTHIA~8 are consistent with charge-independent QCD
dynamics rather than local conservation effects. The combined analysis of CI and CD
observables therefore provides a differential tool to disentangle hard-scattering-induced
correlations, soft particle production, and charge-conservation-driven effects in
small collision systems.

\subsection{Evolution of the near-side correlation peaks}

\begin{figure*} [!htp]
    \centering
    \includegraphics[width=0.42\textwidth]{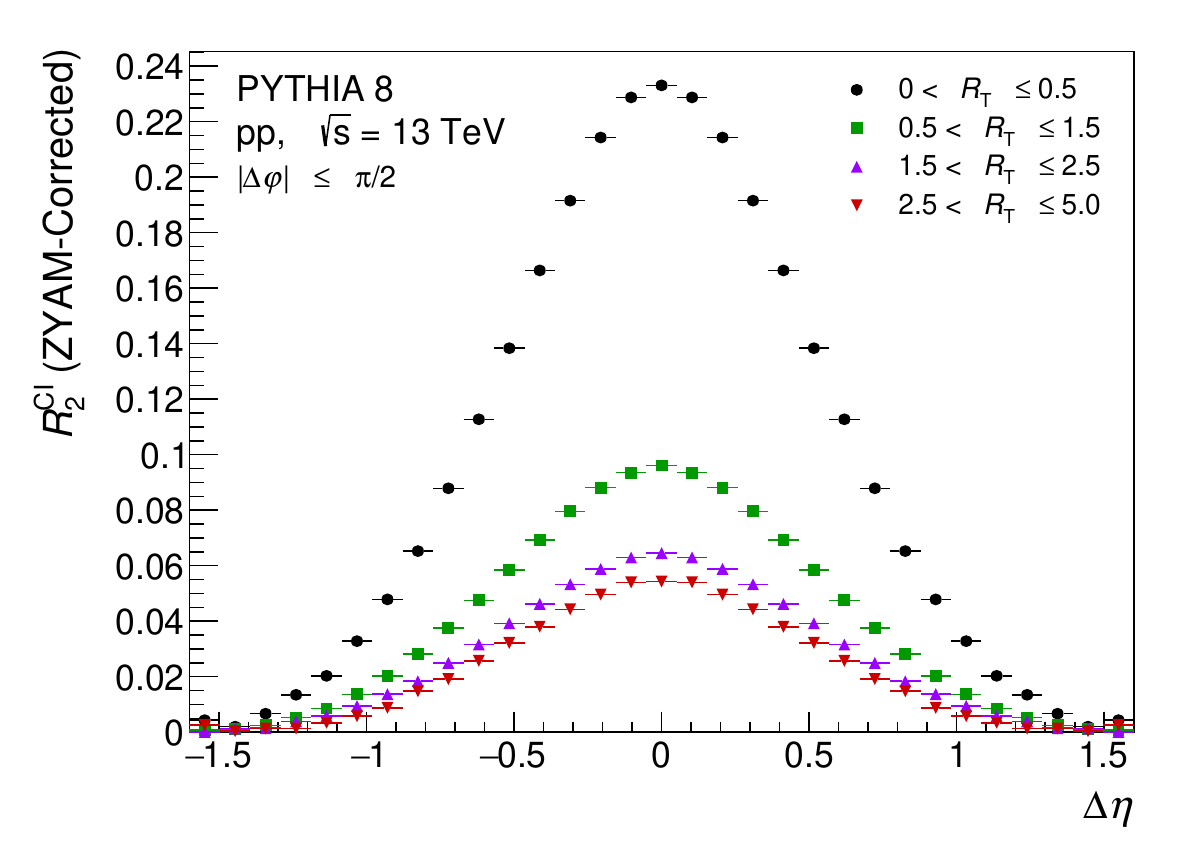}
    \includegraphics[width=0.42\textwidth]{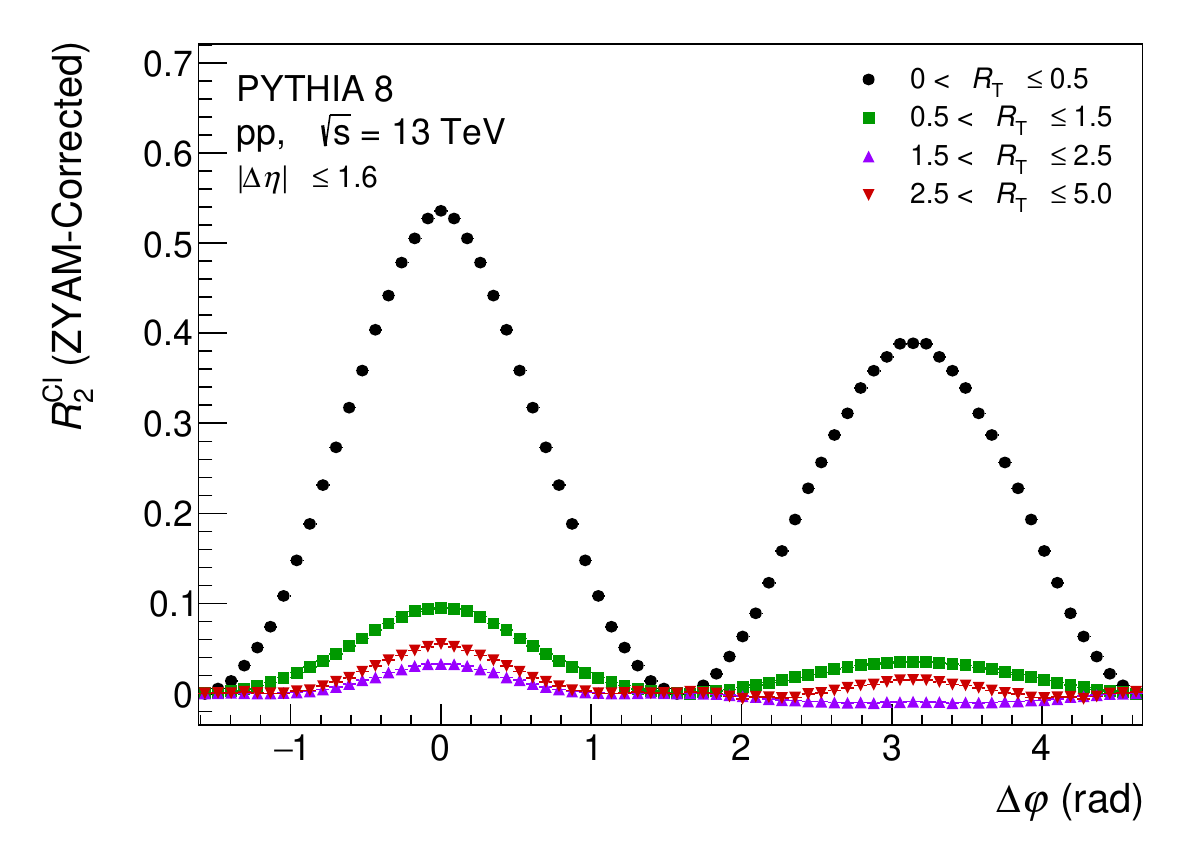}
    
    \includegraphics[width=0.42\textwidth]{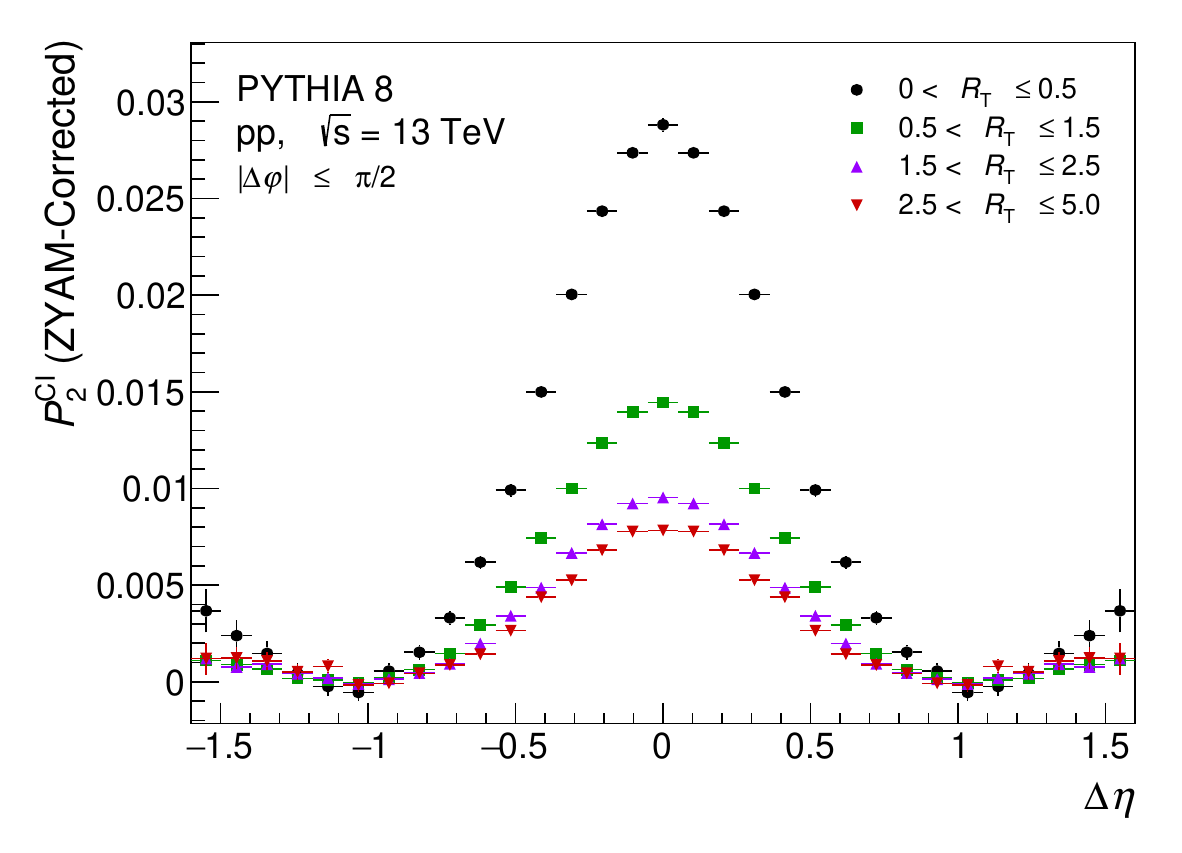}
    \includegraphics[width=0.42\textwidth]{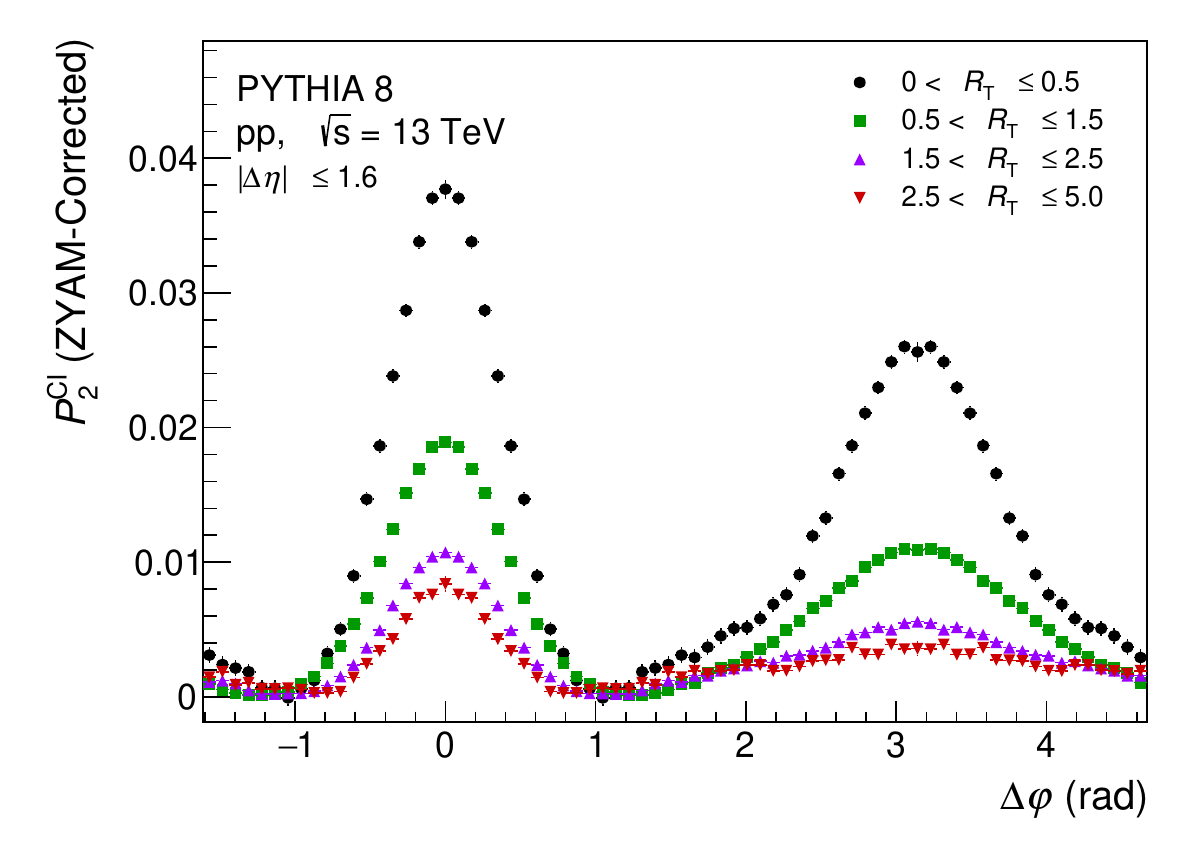}
    \caption{ZYAM-corrected one-dimensional projections of the near-side peaks of the charge-independent correlators $R_{2}^{\mathrm{CI}}$ (top row) and $P_{2}^{\mathrm{CI}}$ (bottom row) 
    along the $\Delta\eta$ (left) and $\Delta\varphi$ (right) directions, shown for four $R_{\rm T}$ event classes in pp collisions at $\sqrt{s}=13$~TeV simulated with PYTHIA~8 (Monash 2013 tune). 
    Projections are taken within $|\Delta\varphi|\leq\pi/2$ and $|\Delta\eta|\leq1.6$, respectively.}
    \label{fig:CIproj_RT4}
\end{figure*}

\begin{figure*} [!htp]
    \centering
    \includegraphics[width=0.42\textwidth]{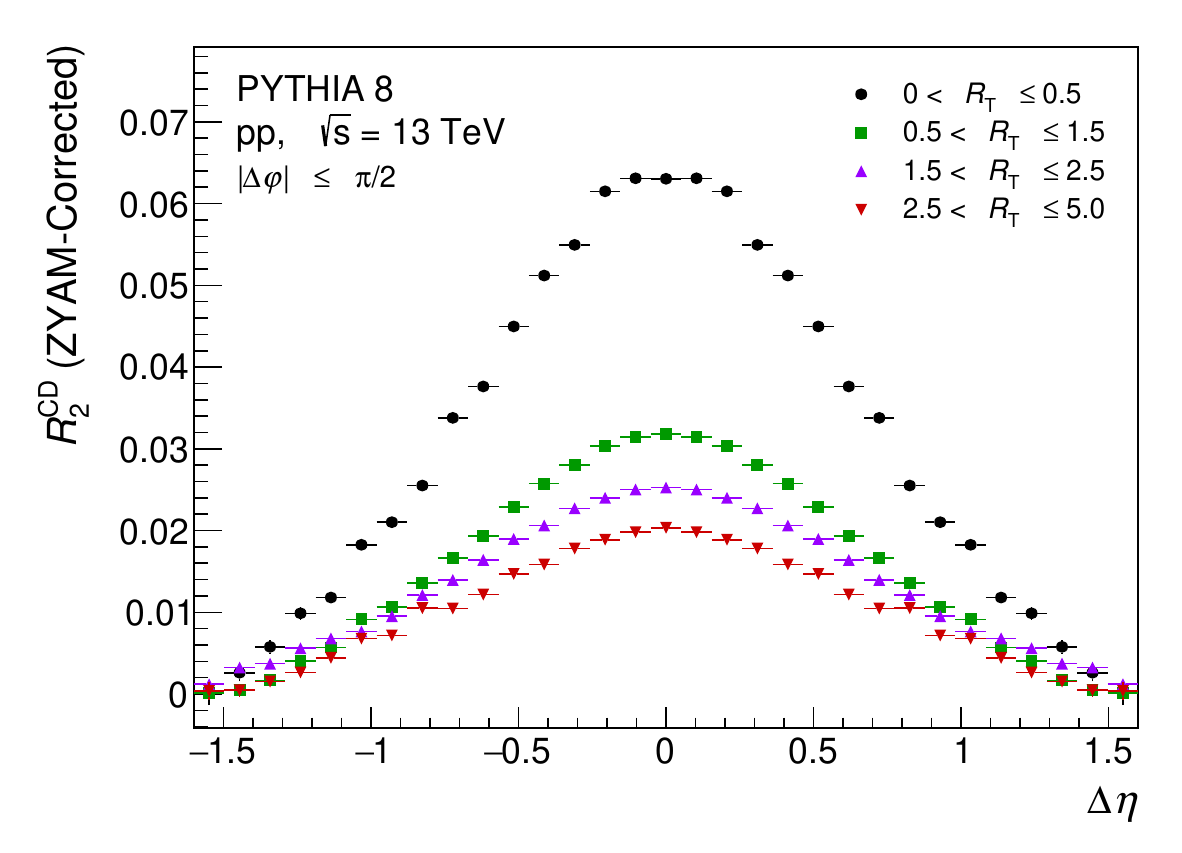}
    \includegraphics[width=0.42\textwidth]{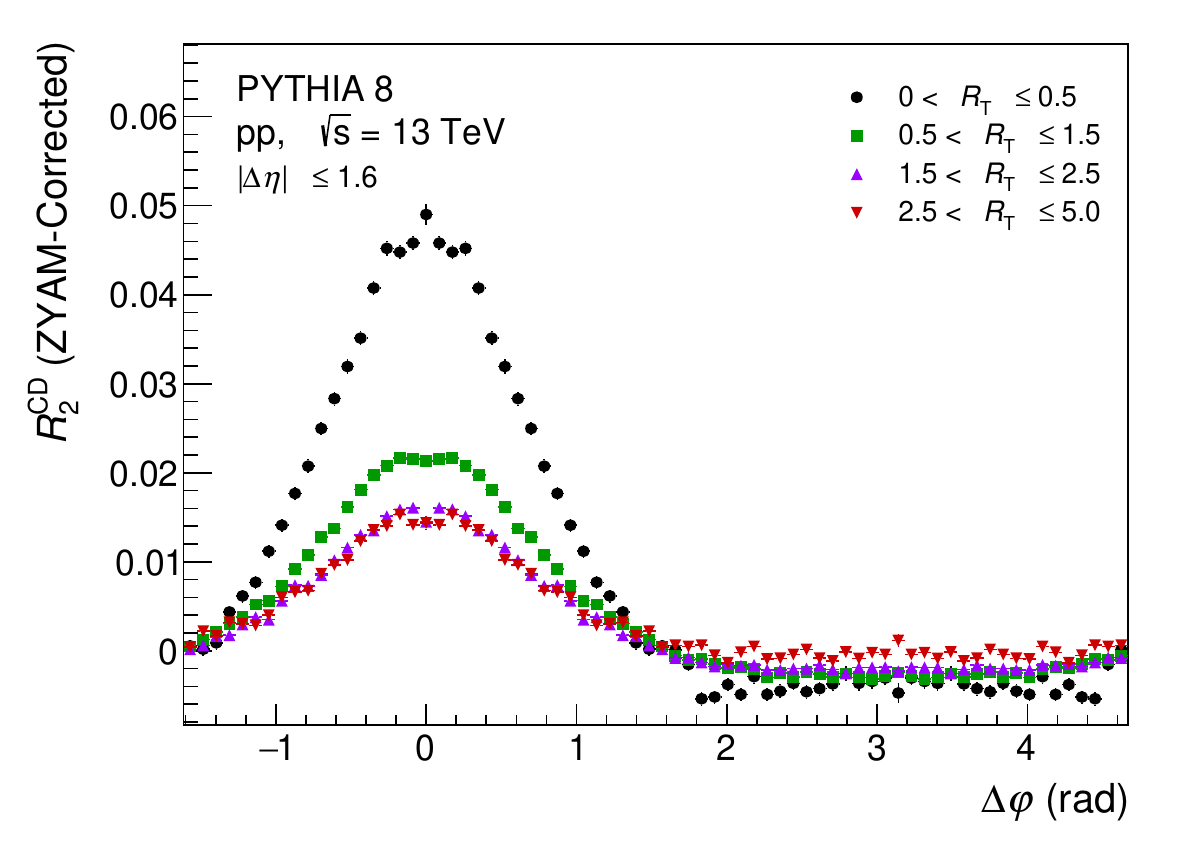}
    
    \includegraphics[width=0.42\textwidth]{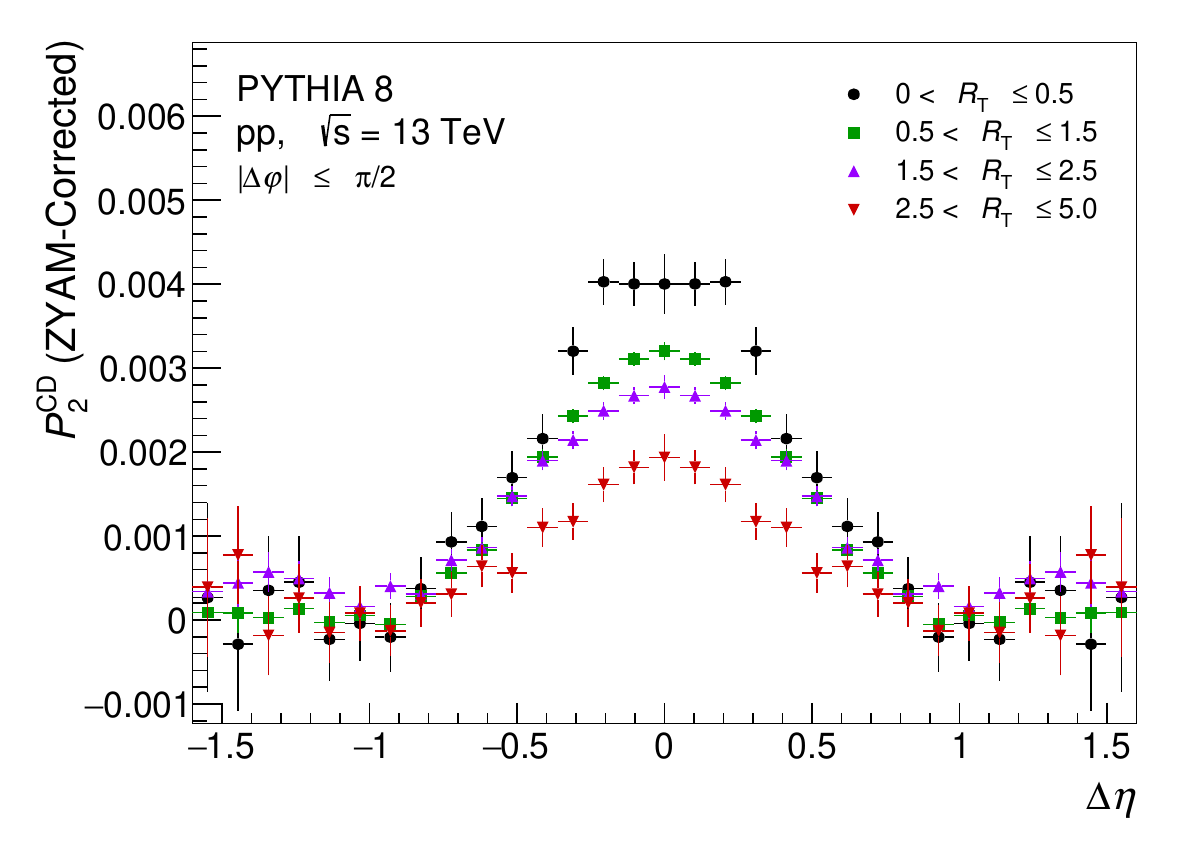}
    \includegraphics[width=0.42\textwidth]{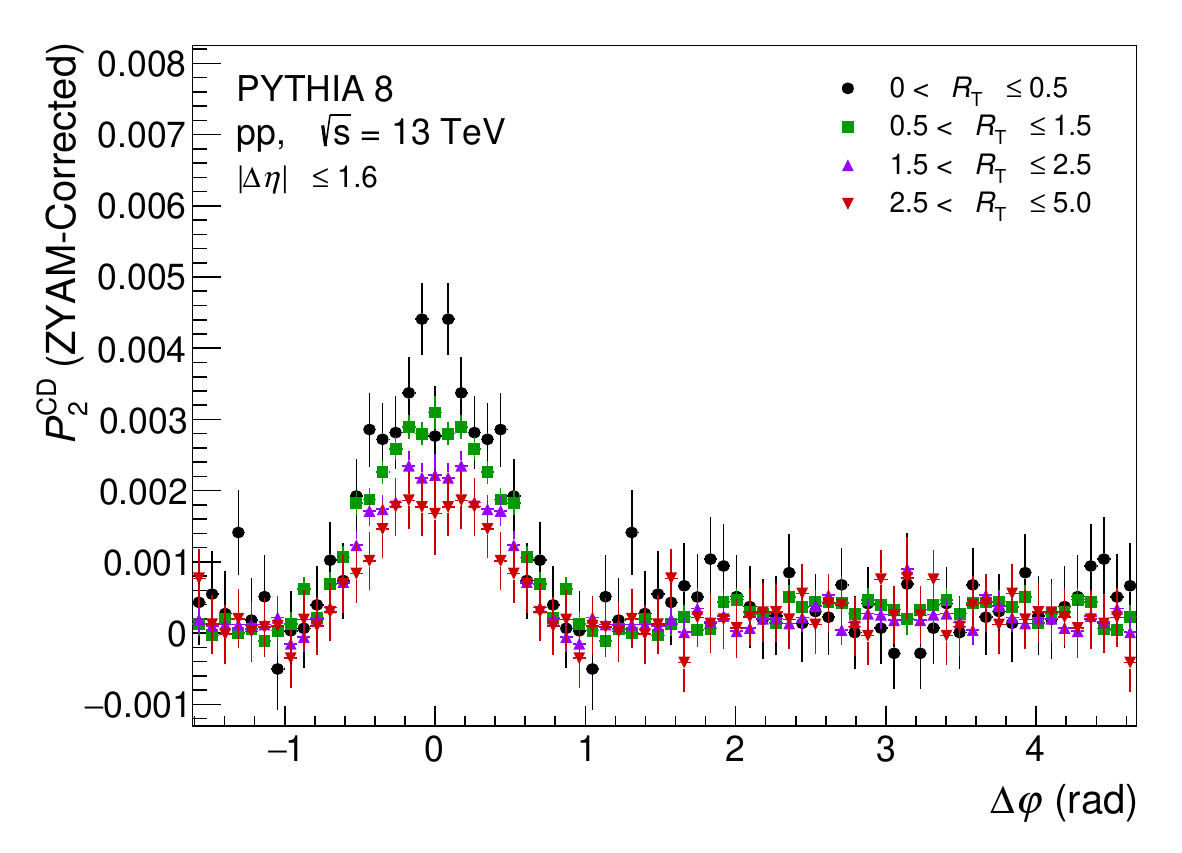}
    
    \caption{ZYAM-corrected one-dimensional projections of the near-side peaks of the charge-dependent correlators $R_{2}^{\mathrm{CD}}$ (top row) and $P_{2}^{\mathrm{CD}}$ (bottom row) 
    along the $\Delta\eta$ (left) and $\Delta\varphi$ (right) directions, shown for four $R_{\rm T}$ event classes in pp collisions at $\sqrt{s}=13$~TeV simulated with PYTHIA~8 (Monash 2013 tune). 
    Projections are taken within $|\Delta\varphi|\leq\pi/2$ and $|\Delta\eta|\leq1.6$, respectively.}
    \label{fig:CDproj_RT4}
\end{figure*}

\begin{figure*} [!htp]
    \centering
    \includegraphics[width=0.42\textwidth]{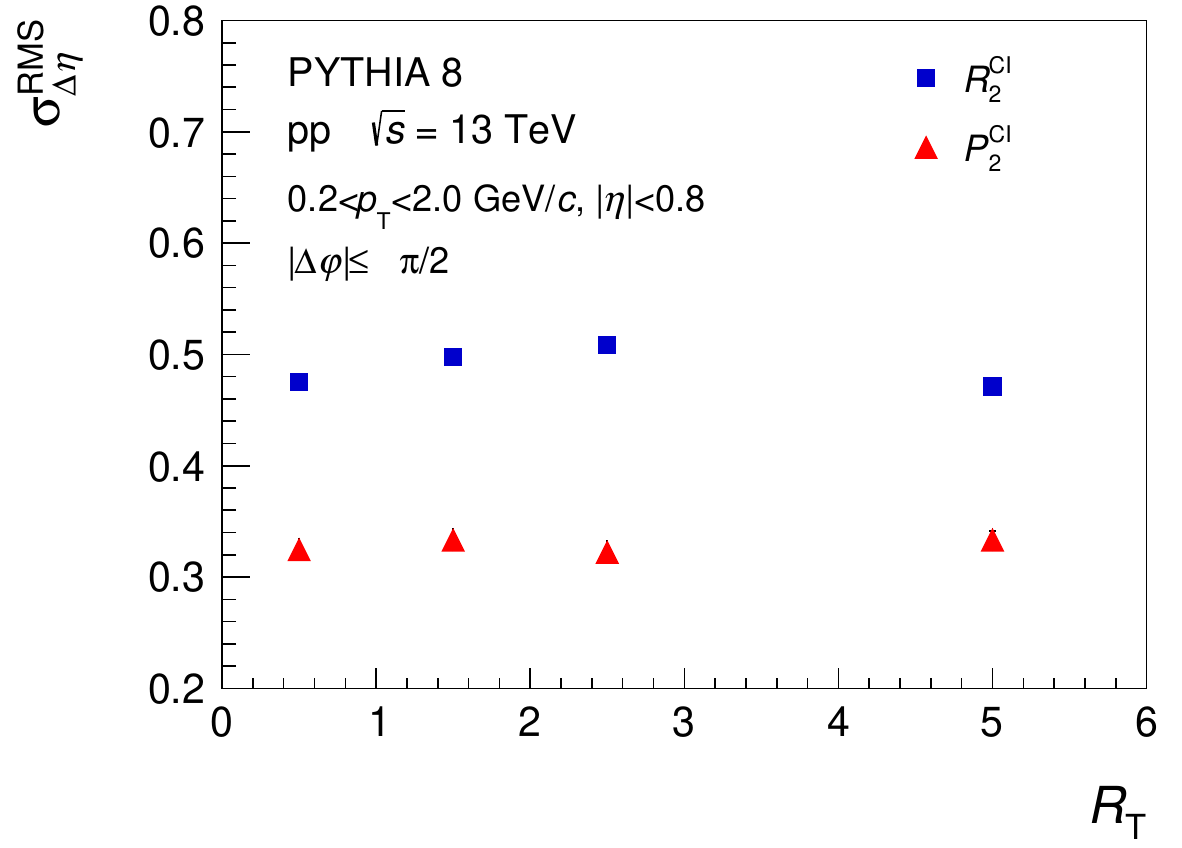}
    \includegraphics[width=0.42\textwidth]{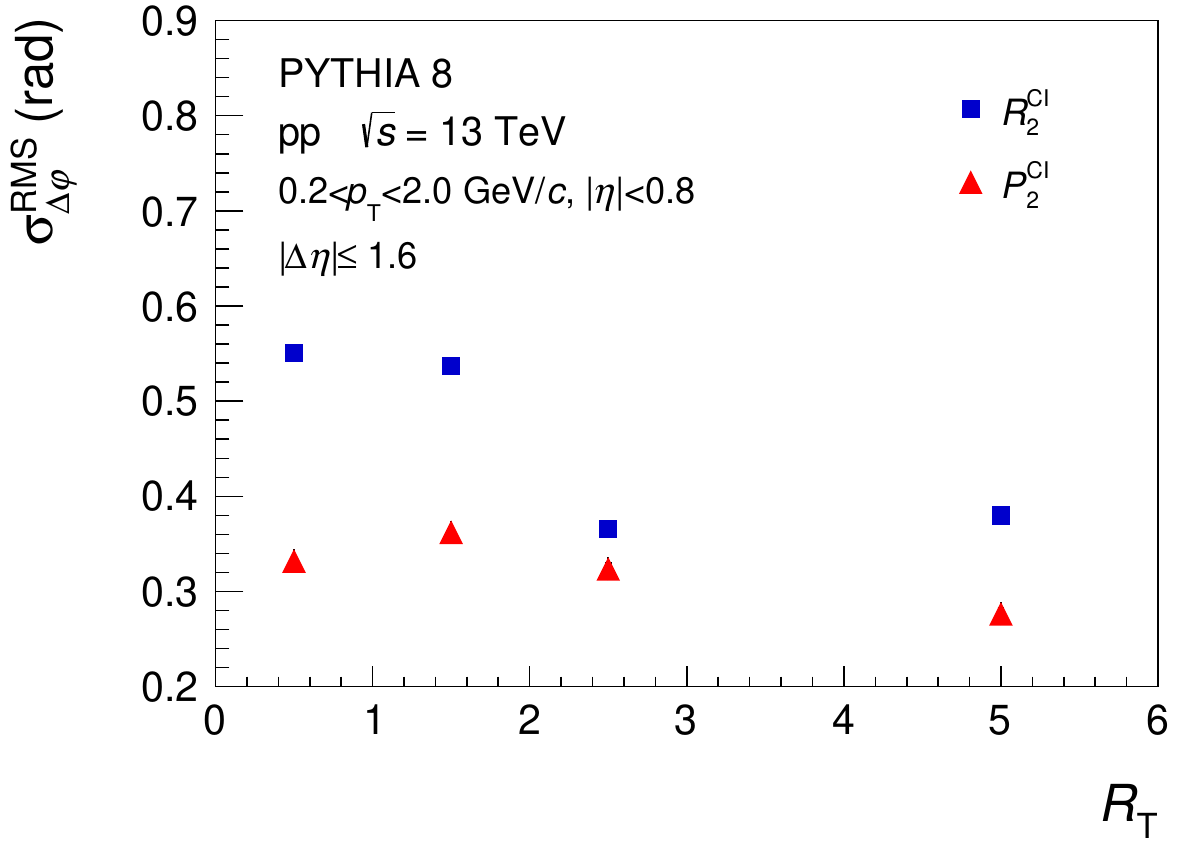}

    \caption{Root-mean-square (RMS) widths of the near-side peaks of the charge-independent correlators $R_{2}^{\mathrm{CI}}$ and $P_{2}^{\mathrm{CI}}$ along the $\Delta\eta$ (left) 
    and $\Delta\varphi$ (right) directions, shown as a function of $R_{\mathrm{T}}$ in pp collisions at $\sqrt{s}=13$~TeV simulated with PYTHIA~8 (Monash 2013 tune). 
    The $\Delta\eta$ widths remain approximately constant across all $R_{\rm T}$ classes, while the $\Delta\varphi$ widths narrow systematically with increasing $R_{\rm T}$, 
    most prominently for $R_{2}^{\mathrm{CI}}$, indicating growing azimuthal coherence driven by MPI and color reconnection. 
    $R_{2}^{\mathrm{CI}}$ is consistently broader than $P_{2}^{\mathrm{CI}}$ in both directions.}
    \label{fig:sigCI_DetaDphi}
\end{figure*}

\begin{figure*} [!htp]
    \centering
    \includegraphics[width=0.42\textwidth]{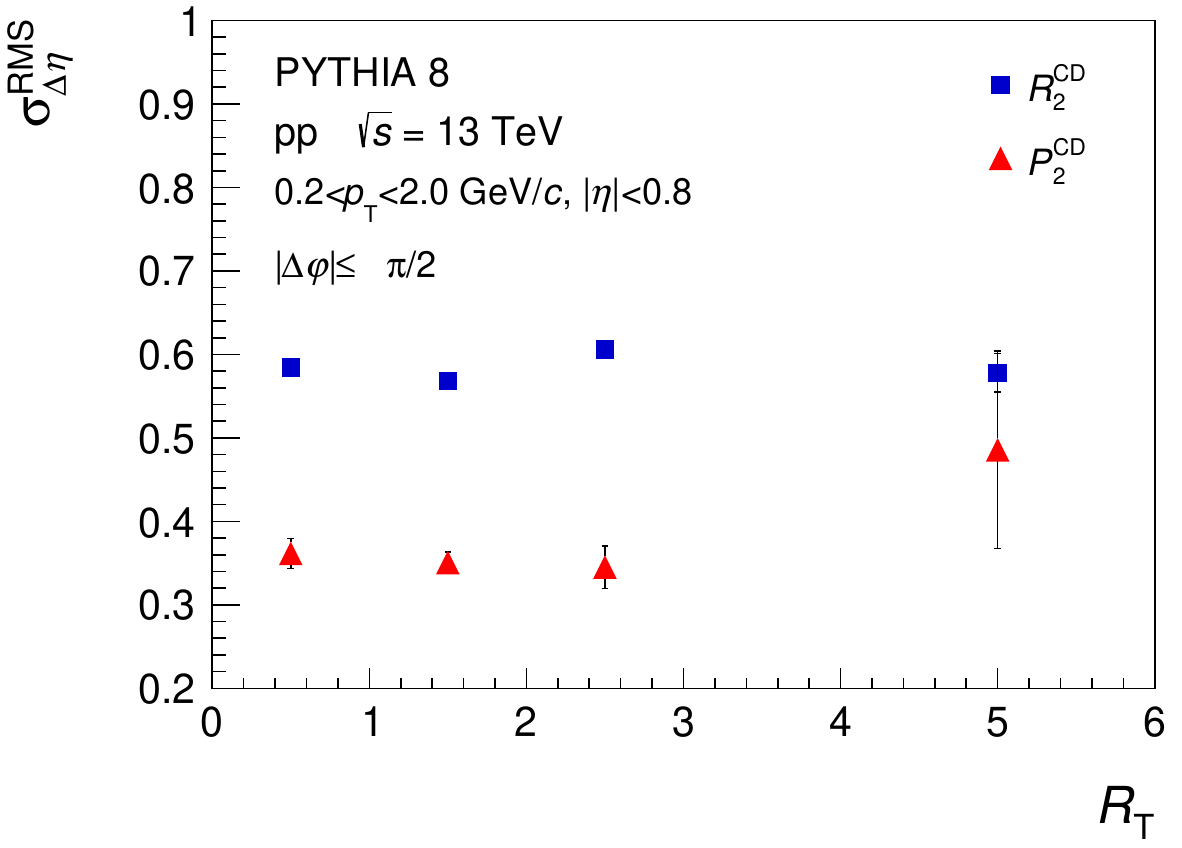}
    \includegraphics[width=0.42\textwidth]{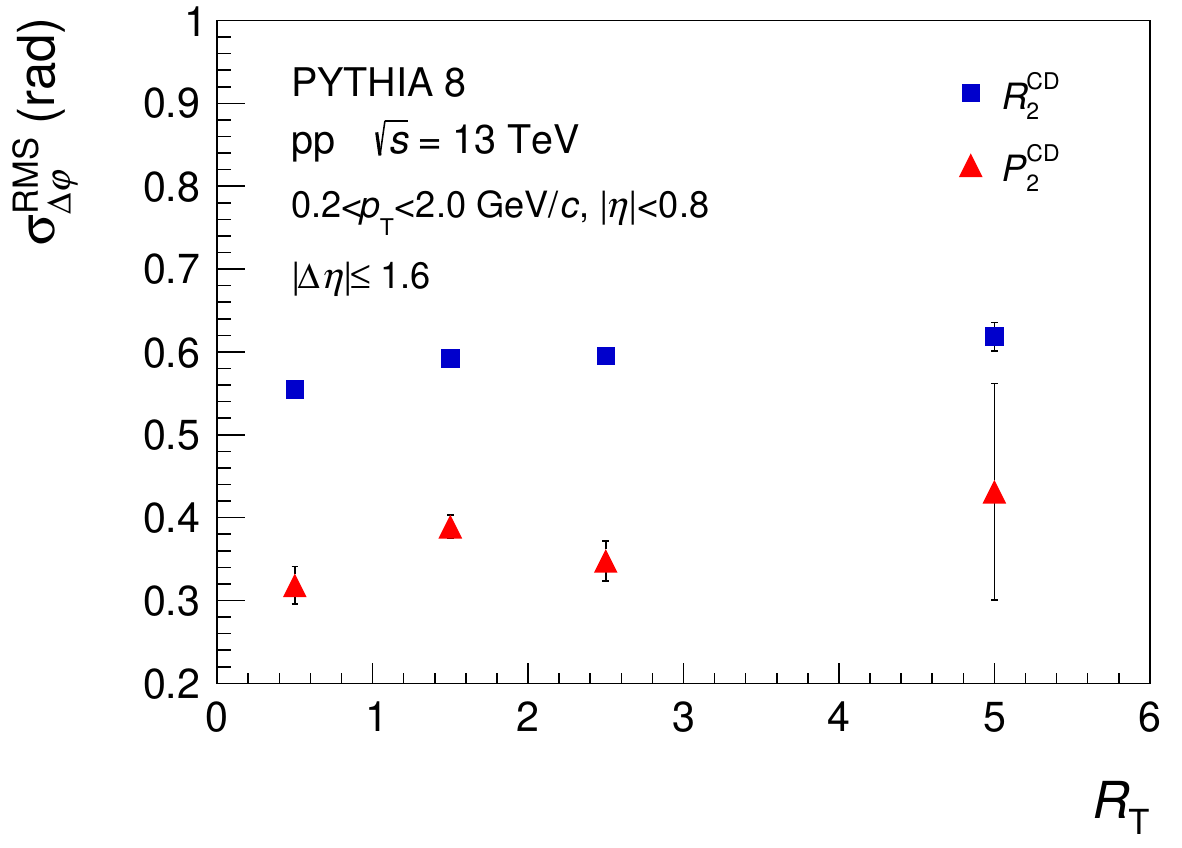}
    
    \caption{Root-mean-square (RMS) widths of the near-side peaks of the charge-dependent correlators $R_{2}^{\mathrm{CD}}$ and $P_{2}^{\mathrm{CD}}$ along the $\Delta\eta$ (left) 
    and $\Delta\varphi$ (right) directions, shown as a function of $R_{\mathrm{T}}$ in pp collisions at $\sqrt{s}=13$~TeV simulated with PYTHIA~8 (Monash 2013 tune). 
    Both widths exhibit a mild but systematic increase with increasing $R_{\rm T}$, indicating a modest broadening of charge-balancing correlations driven by overlapping partonic interactions.}
    \label{fig:sigCD_DetaDphi}
\end{figure*}

Beyond the long-range component, it is equally instructive to examine how the
near-side peak evolves across $R_{\mathrm{T}}$ classes, as the variation of peak
widths provides complementary information on the interplay between
hard-scattering-induced and soft particle-production mechanisms. To this end,
one-dimensional projections of the near-side peak were constructed along the
$\Delta\eta$ and $\Delta\varphi$ directions within $|\Delta\varphi| \leq \pi/2$
and $|\Delta\eta| \leq 1.6$, after subtracting the uncorrelated pedestal via the
ZYAM procedure. The resulting projections for the CI and CD correlators are shown
in Figs.~\ref{fig:CIproj_RT4} and \ref{fig:CDproj_RT4}, and the corresponding
RMS widths are presented in Figs.~\ref{fig:sigCI_DetaDphi} and
\ref{fig:sigCD_DetaDphi}.

For the CI correlators, the near-side widths along $\Delta\eta$ remain approximately
constant across all $R_{\mathrm{T}}$ classes, with $R_{2}^{\mathrm{CI}}$
consistently broader than $P_{2}^{\mathrm{CI}}$. This stability indicates that the
pseudorapidity extent of short-range correlations is largely insensitive to
variations in UE activity, reflecting the predominantly local nature of particle
correlations arising from string fragmentation and parton showering.

In contrast, a systematic narrowing of the near-side peak along $\Delta\varphi$ is
observed with increasing $R_{\mathrm{T}}$, most prominently for
$R_{2}^{\mathrm{CI}}$. This behavior suggests that events with higher UE activity
are accompanied by increased azimuthal coherence in particle emission, which within
the PYTHIA~8 framework may reflect the growing contribution of MPI and color
reconnection, introducing additional azimuthal ordering through color-string
rearrangements. The comparatively weak $R_{\mathrm{T}}$ dependence of
$P_{2}^{\mathrm{CI}}$ is consistent with its greater sensitivity to higher-$p_{\mathrm{T}}$
pairs from hard processes, whose intrinsic collimation is less susceptible to
variations in the soft background.

For the CD correlators, the near-side peaks remain localized in both $\Delta\eta$
and $\Delta\varphi$ across all $R_{\mathrm{T}}$ classes, with no evidence of a
long-range component. The extracted widths show a mild but systematic increase with
increasing $R_{\mathrm{T}}$, indicating a modest broadening of charge-balancing
correlations in events with enhanced UE activity. This behavior is consistent with
predominantly local hadronization mechanisms, where increased soft particle
production and overlapping partonic interactions slightly extend the angular
separation of balancing charges without generating long-range charge-dependent
structures.

Collectively, the evolution of the near-side widths reinforces the picture
established in the preceding sections: the long-range component in
$R_{2}^{\mathrm{CI}}$ at high $R_{\mathrm{T}}$ is consistent with
charge-independent soft QCD mechanisms, whereas short-range CD correlations
remain governed by local conservation effects. The differential $R_{\mathrm{T}}$
dependence of CI and CD peak widths thus provides additional support for the
interpretation that UE activity modifies collective-like azimuthal structures
without altering the fundamentally local nature of charge balancing in PYTHIA~8.

\section{Summary}

This study presents a systematic investigation of two-particle number ($R_{2}$) and
transverse-momentum ($P_{2}$) correlation functions in pp collisions at
$\sqrt{s}=13$~TeV simulated with PYTHIA~8, with events classified by the relative
transverse activity $R_{\mathrm{T}}$. The analysis targets final-state charged
hadrons within $|\eta|<0.8$ and $0.2<p_{\mathrm{T}}<2.0$~GeV/$c$, and investigates
how UE activity shapes two-particle correlation structures across four $R_{\mathrm{T}}$
intervals ranging from jet-dominated to UE-dominated events.

The CI correlators display a near-side peak at $(\Delta\eta,\Delta\varphi)=(0,0)$
and a broad away-side ridge at $\Delta\varphi\approx\pi$ across all $R_{\mathrm{T}}$
classes, with the jet-dominated class ($0<R_{\mathrm{T}}\leq 0.5$) exhibiting the
most pronounced structures including a characteristic $\Delta\eta$-elongated
near-side ridge in $R_{2}^{\mathrm{CI}}$. With increasing $R_{\mathrm{T}}$, both
peaks progressively weaken as jet-induced correlations are diluted by growing UE
activity. Across all classes, $P_{2}^{\mathrm{CI}}$ exhibits systematically narrower
near-side peaks than $R_{2}^{\mathrm{CI}}$, consistent with prior ALICE measurements
and attributable to its suppression of soft pairs near $\langle p_{\mathrm{T}}\rangle$.
Notably, a long-range near-side component emerges in $R_{2}^{\mathrm{CI}}$
exclusively in the highest $R_{\mathrm{T}}$ class ($2.5<R_{\mathrm{T}}\leq 5.0$),
confirmed through ZYAM-corrected $\Delta\varphi$ projections and a
pseudorapidity-slice analysis that rules out a residual jet contribution. This
structure is consistent with azimuthal coherence generated by MPI and CR within
PYTHIA~8, rather than hydrodynamic collectivity. In contrast, $P_{2}^{\mathrm{CI}}$
shows no significant long-range component in any $R_{\mathrm{T}}$ class, reflecting
its reduced sensitivity to soft, isotropic particle production.

The CD correlators exhibit single, localized near-side peaks with no long-range
component in any $R_{\mathrm{T}}$ class, consistent with local charge balancing
through string fragmentation, and $P_{2}^{\mathrm{CD}}$ is systematically narrower
than $R_{2}^{\mathrm{CD}}$. Both CD peak magnitudes decrease with increasing
$R_{\mathrm{T}}$, reflecting dilution by uncorrelated soft particles from MPI. The
absence of a long-range structure in the CD correlators, even in the UE-dominated
class, indicates that the ridge in $R_{2}^{\mathrm{CI}}$ arises from
charge-independent QCD mechanisms rather than local charge-conservation effects.

Near-side peak widths offer complementary support for this picture: the CI widths
along $\Delta\eta$ remain stable with $R_{\mathrm{T}}$, while the corresponding
$\Delta\varphi$ widths narrow progressively, most prominently for $R_{2}^{\mathrm{CI}}$,
consistent with growing azimuthal coherence from MPI and CR. The CD widths, by
contrast, show only mild broadening with $R_{\mathrm{T}}$ and develop no long-range
structure. Collectively, these results demonstrate that $R_{\mathrm{T}}$ serves as
a powerful differential event classifier for isolating and characterizing
collectivity-like structures in pp collisions. The combined CI and CD analysis of
$R_{2}$ and $P_{2}$ across $R_{\mathrm{T}}$ classes establishes a well-defined
non-hydrodynamic baseline for interpreting such signatures in LHC measurements,
demonstrating that long-range azimuthal structures in small systems can arise from
microscopic QCD dynamics without invoking hydrodynamic collectivity.

\section{Acknowledgements}
This work was supported by the Department of Science and Technology (DST) and the
University Grants Commission (UGC), India.

\bibliography{myref}

\end{document}